\documentclass{webofc}
\usepackage[varg]{txfonts}   
\begin{document}

\title{The CMS monitoring infrastructure and applications}

\author{
    \firstname{Christian} \lastname{Ariza-Porras}\inst{1}\fnsep\thanks{\email{christian.ariza.porras@cern.ch}}
    \and
    \firstname{Valentin} \lastname{Kuznetsov}\inst{2}\fnsep\thanks{\email{vkuznet@protonmail.com}}
    \and
    \firstname{Federica} \lastname{Legger}\inst{3}\fnsep\thanks{\email{federica.legger@cern.ch}}
}
\institute{
    CERN, Geneva, Switzerland
    \and
    Cornell University, Ithaca, NY, USA 14850
    \and
    Istituto Nazionale di Fisica Nucleare, via Pietro Giuria 1, Torino, Italy
}



\abstract{
The globally distributed computing infrastructure required to cope with the multi-petabytes datasets produced by the Compact Muon Solenoid (CMS) experiment at the Large Hadron Collider (LHC) at CERN comprises several subsystems, such as workload management,  data management, data transfers, and submission of users' and centrally managed production requests. The performance and status of all subsystems must be constantly monitored to guarantee the efficient operation of the whole infrastructure. Moreover, key metrics need to be tracked to evaluate and study the system performance over time.
The CMS monitoring architecture allows both real-time and historical monitoring of a variety of data sources and is based on scalable and open source solutions tailored to satisfy the experiment's monitoring needs. We present the monitoring data flow and software architecture for the CMS distributed computing applications. We discuss the challenges, components, current achievements, and future developments of the CMS monitoring infrastructure. 
}
\maketitle

\section{Introduction}
\label{intro}


Data from the CMS experiment \cite{cms} at the LHC are stored and processed in a tiered distributed computing infrastructure. More than one hundred
computing centers worldwide, connected through a set of grid services responsible for storage, computing, and connectivity
are used to process LHC data and produce Monte Carlo simulated events of relevant physics processes. A recent overview of the computing model of the LHC experiments can be found in \cite{wlcg}. 

The CMS distributed computing infrastructure includes central services and middleware that handle authentication, workload management, and data management. The workload management system executes payloads in compute nodes provisioned through GlideinWMS \cite{glidein} and made available as execution slots in a Vanilla Universe HTCondor pool \cite{condor}. HTCondor jobs are submitted via specific workload management tools: WMAgent for central data processing and Monte Carlo production jobs, and CRAB for user jobs \cite{crab}. The data management system is modular, and includes several components:  PhEDEx, the data transfer and location system; DBS, the Data Bookkeeping Service, a metadata catalog; and DAS, the Data Aggregation Service designed to aggregate views and provide them to users and services \cite{DM}. Data from these services are available to CMS collaborators through a web suite of services known as CMSWEB. 

Until recently, most services were monitored through custom tools and web applications, and logging information was scattered over several sources and mostly accessible only by experts. The maintenance and operation of the monitoring applications were becoming increasingly time-consuming and complicated due to the high turnover of experts in the computing community. 

Nowadays, several solutions are available to gather, store, and process large amounts of data such as those produced by monitoring and logging services of computing applications. Many technologies are available under an open-source licence, and are developed and supported by large software companies and communities. 

During the last two years, the CMS computing community has been gradually abandoning in-house solutions in favour of the adoption of widely-used technologies based on open-source, scalable, and non-SQL tools, such as Hadoop \cite{HDFS}, InfluxDB \cite{InfluxDB}, and ElasticSearch \cite{ES}. These services are available through MONIT \cite{monit-paper}, a central  monitoring infrastructure provided by the CERN IT department, and allow for the easy deployment of monitoring and accounting applications using visualisation tools such as Kibana \cite{kibana} and Grafana \cite{grafana}. Grafana allows users to automatically raise alarms and send notifications when anomalous conditions in the monitoring data are met. Data sources from different subsystems can be used to build complex monitoring workflows and predictive analytics, and performance studies.  To complement the monitoring infrastructure provided by the CERN IT department, the CMS monitoring group set up and runs additional monitoring applications based on technologies such as VictoriaMetrics \cite{VM}, NATS \cite{NATS}, and Prometheus \cite{prometheus}.
 
In the following sections, we describe MONIT, the monitoring infrastructure at CERN, the organisation of CMS monitoring applications based on MONIT and CMS own infrastructure, and future developments.

\section{The CERN MONIT infrastructure}
\label{monit}

\begin{figure*}[t]
\centering
\includegraphics[scale=0.28]{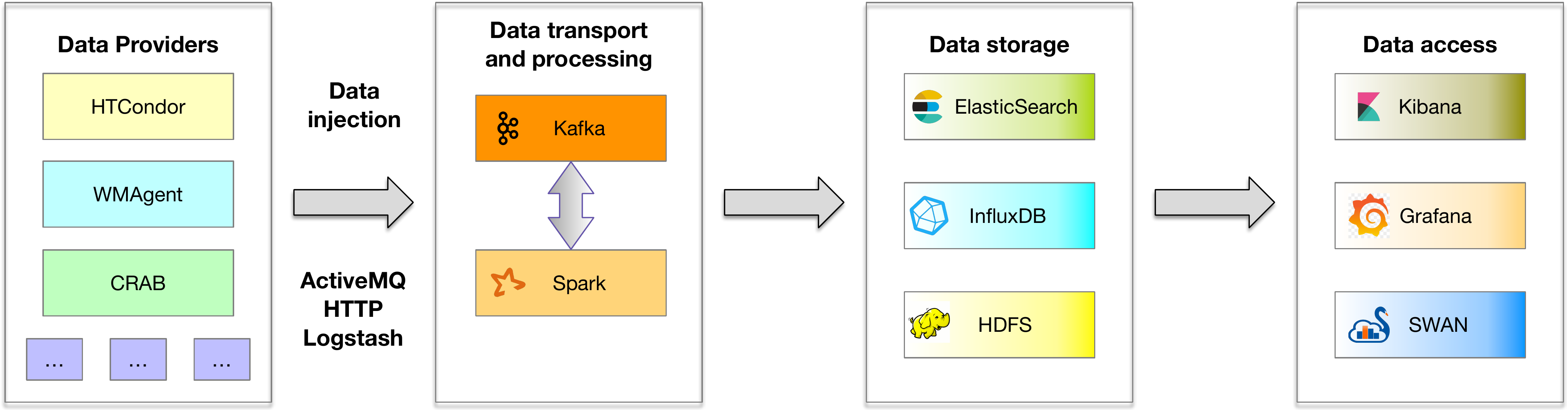}

\caption{The data flow in the CERN MONIT infrastructure. 
The various steps of a typical monitoring workflow (data injection, transport, processing, storage, and access) are shown from left to right.}
\label{cern-monit}
\end{figure*}

The CERN IT department offers a variety of monitoring services through the MONIT infrastructure. A typical monitoring workflow consists of the following steps (see also Fig. \ref{cern-monit}):
\begin{itemize}
    \item \textbf{Data injection:} data is pushed into the MONIT infrastructure through ActiveMQ \cite{amq} messaging. As an alternative, data providers inside the CERN network boundaries may use an HTTP endpoint. Log files may be directly injected using  Logstash \cite{logstash}.
    \item \textbf{Data transport and processing:} Data are streamed into MONIT using Apache Kafka \cite{kafka}.
     Apache Spark \cite{spark} and Apache Flume \cite{Flume} are used for further data processing. Data may be enriched with additional information or data from several sources can be aggregated as needed.
    \item \textbf{Data storage:} Three storage technologies are available, depending on the needs of the particular use case: data volume, schema, and retention policy. ElasticSearch (ES) is used to store semi-structured data for a short period of time (up to one month). InfluxDB is used to store time-series structured, aggregated data for a longer period of time (up to five years, depending on the granularity). The Hadoop Distributed File System (HDFS) is used for long-term (currently unlimited in time) storage of data in a variety of formats (Apache Avro \cite{avro}, Apache Parquet \cite{parquet}, JSON \cite{json}, plain text).
    \item \textbf{Data access:} For data access we rely on the stack ES/Kibana, Grafana, and the SWAN \cite{swan} service at CERN. Kibana is used for data exploration and visualisation of ES data sources. Grafana is a visualisation tool that can read data from several sources such as ES, InfluxDB, Prometheus, Graphite \cite{graphite}, Open TSDB \cite{openTSDB}, and MySQL \cite{mysql}. Data sources can also be accessed through the Grafana proxy. The Apache Spark framework is used to process data on HDFS. The SWAN service provides access to the CERN Hadoop clusters through a JupyterHub interface \cite{jupyter}.
\end{itemize}

The ActiveMQ servers and the SWAN service are not strictly part of the MONIT infrastructure, but are offered and managed by the CERN IT department.

\section{The CMS monitoring infrastructure}
\label{cmsmonit}

In this section we provide a comprehensive overview of various components of the CMS monitoring infrastructure. In Section \ref{cms-monit-use} we describe how CMS is building monitoring applications based on the  MONIT infrastructure. In Section \ref{cms-monit-addons} we provide details of the CMS specific monitoring infrastructure, and we discuss the CMS monitoring Kubernetes (k8s) \cite{k8s} clusters in Section \ref{cmsmonit-k8s}.

\subsection{CMS usage of the MONIT infrastructure}
\label{cms-monit-use}

A variety of CMS data producers regularly inject data into the MONIT infrastructure. There are currently more than twenty-five active CMS monitoring workflows, covering a large variety of systems: HTCondor job parameters, GlideinWMS submission infrastructure, data transfers and access patterns, CRAB and WMAgent tasks, CMSWEB services. With reference to the data flow outlined in Section \ref{monit}, we describe in the following paragraphs how CMS is using the MONIT infrastructure.

\subsubsection{Data injection}
\label{data-injection}

 Most of CMS monitoring data producers use the ActiveMQ endpoints — with a few exceptions for producers inside the CERN internal network which use an HTTP REST endpoint — to inject data in JSON format. The CERN MONIT infrastructure consumes the data via a Kafka pipeline and redirects them to the ES, InfluxDB, and HDFS data sinks. We do not impose a specific schema on injected documents, but the schema should be consistent over time. The document schema and daily data volume are agreed a priory for each CMS data producer with the CERN MONIT team, which then allocates and controls the  resources to consume the data in their infrastructure.

\subsubsection{Data storage}
\label{data-storage}   
    
The CMS data producers exploit all the provided storage options depending on their particular use case. Data in ES are stored for the last 30-40 days depending on the data source. In addition to the raw data indexes, we take advantage of the document indexing features of ES to create  purpose-specific indexes which incorporate some additional logic. For example, we compute and aggregate over HTCondor jobs that have been retried several times, or over temporary job statuses. InfluxDB is used to store aggregated structured data as time series for a limited set of tags and fields. This is used to build historical views of key performance metrics.


\subsubsection{Data access}
\label{data-access} 

Access to the data stored in MONIT is provided through visualization tools such as Grafana and Kibana for data in InfluxDB and ES, and Apache spark jobs (either through the SWAN service or standalone scripts) for data in HDFS. Kibana is mainly used for interactive data exploration of the ES data sources. The search capabilities of ES enable our users to quickly create ad-hoc queries and visualizations. Most CMS official dashboards are implemented in Grafana. Additionally, users can create their own personalised views.
    
Data stored in HDFS are typically accessed using Spark-based workflows, and are used to create dedicated views over long time periods (for example, to regularly produce yearly data popularity plots), or for complex queries and visualizations that can not be implemented in Grafana or Kibana.

\subsubsection{Experiences with the MONIT infrastructure}
\label{cms-monit-experience}

\begin{figure*}[!ht]
\centering
\includegraphics[width=0.95\textwidth,trim={1cm 1cm 1cm 1cm},clip]{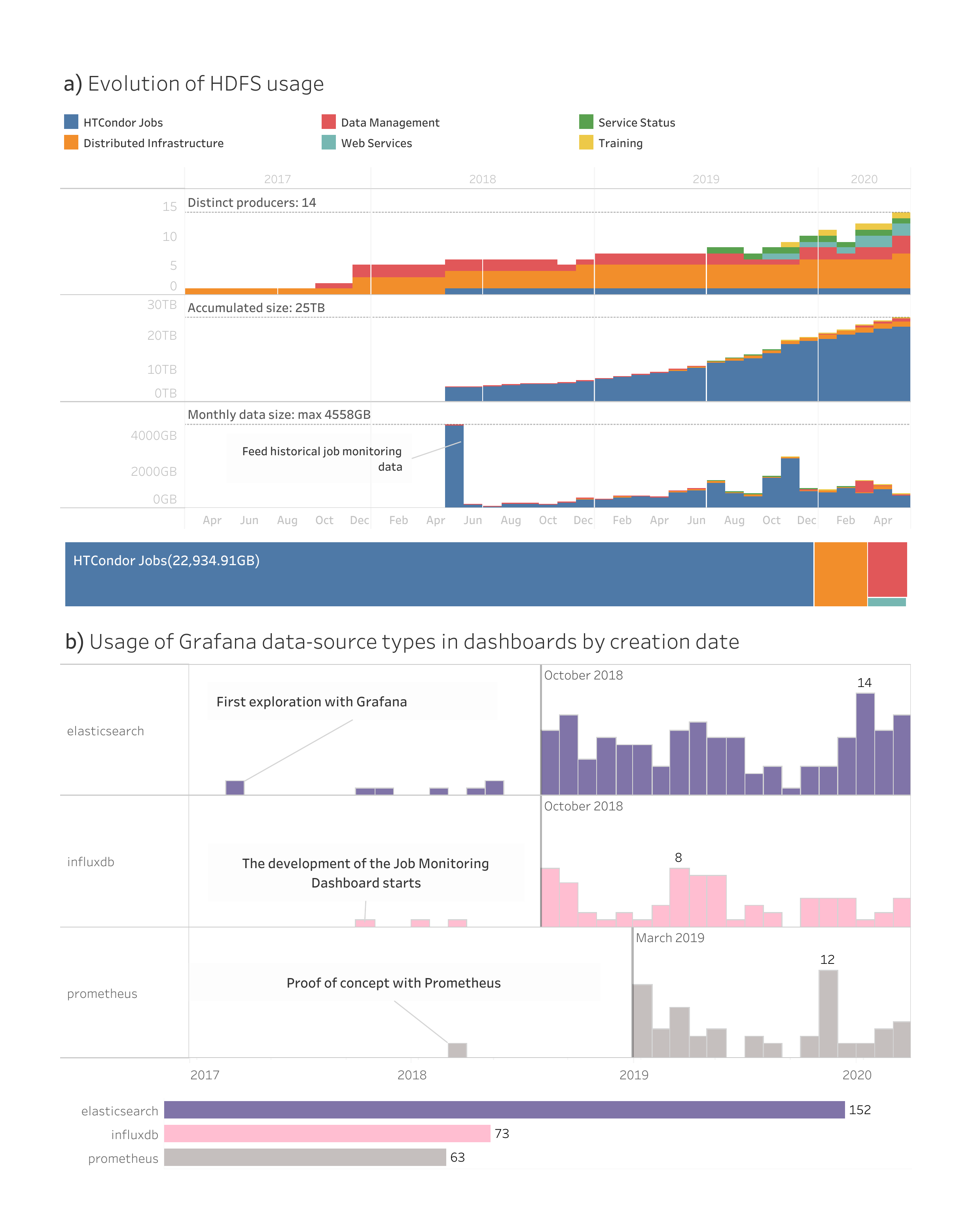}
\caption{Time evolution distributions showing the development and adoption of the CMS Monitoring infrastructure. a) From top to bottom, the evolution of the total number of CMS data producers, the data stored in HDFS, and the monthly data volume. b) The number of Grafana dashboards using ES, InfluxDB, and Prometheus as data sources, ordered by date of creation.}
\label{cms-monit-evolution}
\end{figure*}

In the past two years several custom CMS monitoring applications have been ported to the MONIT infrastructure. The first tests with the new infrastructure started in 2017, and the migration process formally began in October 2018. Since then the number of data sources, stored data volume, and usage has steadily grown (see Fig \ref{cms-monit-evolution}). 

HTCondor job monitoring is one of the most important views, since it provides an overview of the performances of the CMS distributed computing infrastructure, and was one of the first to be migrated. Data from the HTCondor pool are collected, processed, and injected into MONIT at regular intervals (every twelve minutes) by a service, implemented in python, called spider. The spider sends data to MONIT through the ActiveMQ endpoint. The HTCondor data are finally stored in ES, InfluxDB, and HDFS. 

The complete migration took about one year through various steps: first an evaluation phase, where all use cases were gathered and requirements for the new application drafted, then the actual development phase, where the complete data flow was implemented, then a final evaluation phase, where the new data flow was validated against the old one, first from CMS monitoring experts, and then from the wider CMS community. The migration required the coordination and interplay of several teams: the MONIT team, the developers of the CMS monitoring group, and the CMS stakeholders (the data providers and consumers). In particular, it was crucial to define the schema of the data in  a way that all user requirements can be satisfied using the visualization options of Grafana or Kibana. 

A major problem we encountered was the lack of schema validation in the MONIT infrastructure which may lead to some documents being rejected by ES, if the same field has different types in different documents. To avoid this problem, we introduced schema validation of the injected documents through the CMS client tools, and we are gradually enforcing it for all CMS data providers. We also noticed that some attributes are reserved by ES (such as version, timestamp, uuid), and those should be avoided.
 
In InfluxDB a limited set of aggregated tags are stored. The aggregations are computed every twelve minutes and stored for one week. Additional aggregations, e.g. one-hour, one-day, seven-days and thirty-days bins are stored for five years. We currently store eighteen tags, with a series cardinality (unique combination of tags values) of almost nine million. The performance of InfluxDB is affected by the cardinality of the stored time series, and that imposes strong limits on the number and type of tags that we can store. The MONIT team is currently evaluating a future retirement of the InfluxDB workflow for this use case, and its replacement with a dedicated ES index to store the time aggregated data.

With the experience gained in the migration of the HTCondor job monitoring application, more custom workflows have been moved to MONIT in the past months: data popularity and data access, and the Site Status Board, a monitoring application that gathers metrics about performances and status of all CMS computing sites. Several analytics workflows have been added to exploit the wealth of data stored in HDFS, which allows in-depth study of several performance metrics, such as the evolution in time of CPU efficiencies for HTCondor jobs, or resource requests for memory, CPU's, storage. 

\subsection{CMS additional components}
\label{cms-monit-addons}

The MONIT infrastructure does not cover all CMS monitoring requirements, for example the need for quasi-real time monitoring of CMS computing nodes, complex HTCondor job workflows, applications, and services. To fill this gap, we deployed additional monitoring services based on tools such as Prometheus, VictoriaMetrics and AlertManager \cite{PROM-AM}. The choice of these tools was based on their wide adoption, versatile performance, and solid reputation in the IT world. The CMS monitoring infrastructure includes the following  components (see also Fig. \ref{cmsmonitinfra}):

\begin{figure*}[!t]
\centering
\includegraphics[scale=0.35]{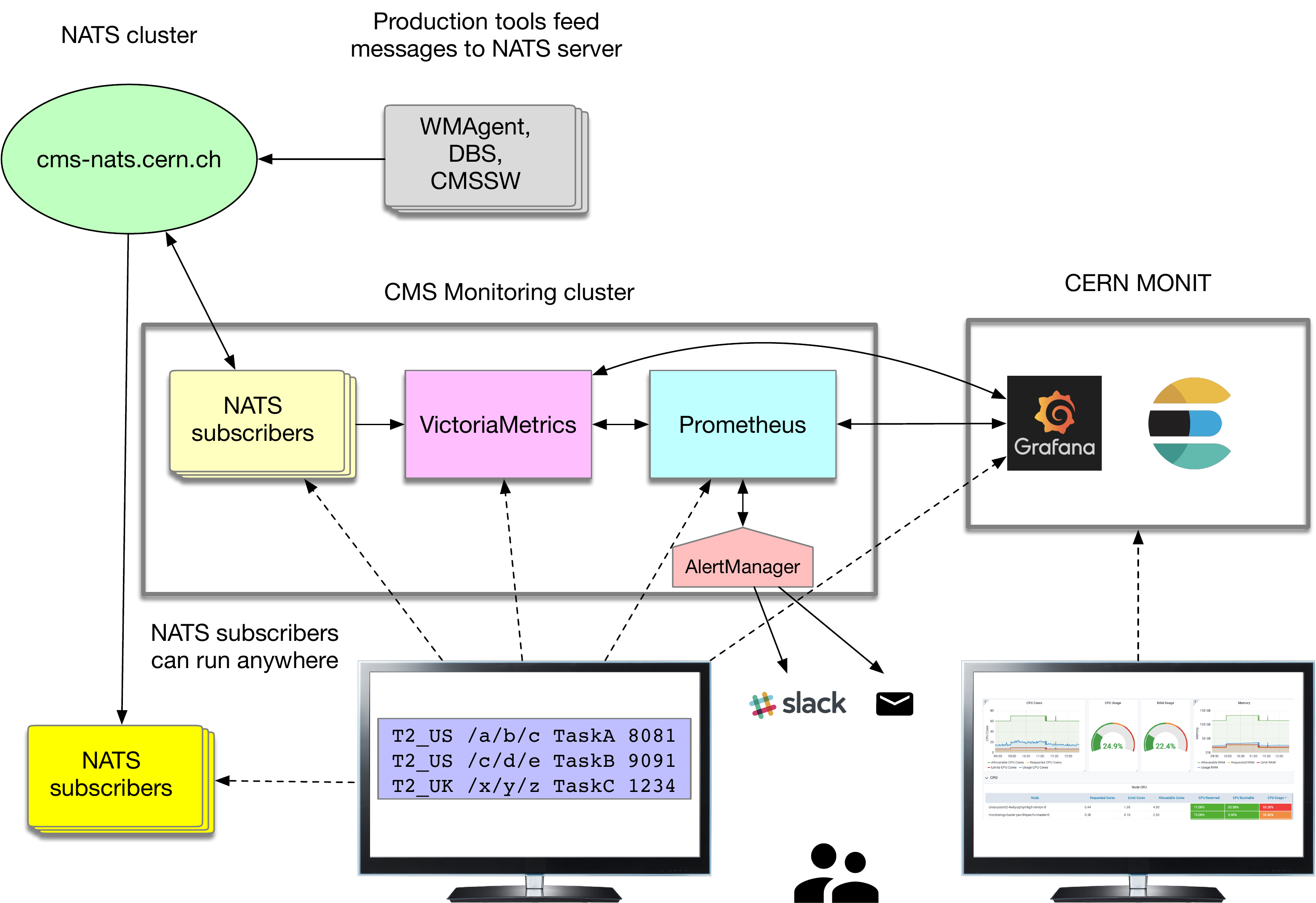}
\caption{The architectural diagram of the CMS monitoring infrastructure. The MONIT infrastructure (described
in Section \ref{cern-monit}) is shown in the right box, and the components maintained and provided by CMS (described in Section \ref{cms-monit-addons}) are displayed in the
left box.
}
\label{cmsmonitinfra}
\end{figure*}

\begin{itemize}
\item the Prometheus service used to monitor various CMS services, for example, CMS web applications;
\item the Logstash service used to parse and send the CMSWEB logs to the MONIT timber service;
\item the NATS (Neural Autonomic Transport System) service to provide real-time messaging to monitor the status of CMS production workflows and campaigns;
\item Pushgateway, a service provided by Prometheus that allows to push metrics from jobs which cannot be scraped, such as ephemeral and batch jobs;
\item VictoriaMetrics, a fast and efficient time series DB,  used as long-term storage for Prometheus. It allows
the integration of NATS messages into Prometheus and Grafana;
\item the AlertManager to handle alerts for Prometheus metrics.
\end{itemize}
In addition we developed and maintain a set of scripts and services for several purposes:  

\begin{itemize}
\item the spider service described in Section \ref{cms-monit-experience} to collect the HTCondor job parameters and push them into MONIT;
\item the CMSSpark framework \cite{CMSSpark} that provides common access to a variety of CMS data sources on HDFS. CMSSpark is used by several workflows, for example to collect and aggregate the data used for the  data popularity views;
\item various CMS databases, such as DBS, are regularly dumped in HDFS by a set of Apache Sqoop \cite{sqoop} jobs.
\end{itemize}

Prometheus and VictoriaMetrics are the main data sources. We store metrics in Prometheus for fifteen days, and up to one month in VictoriaMetrics. We plan to increase further the time retention policy for data in VictoriaMetrics. 

The maintenance of such a complex infrastructure represents certain challenges, such as service deployment, version control, and resource utilization. We gradually migrated the monitoring services described above to a k8s infrastructure to fully  leverage its ease of deployment, the dynamic scalability, and minimal maintenance costs. The CMS k8s monitoring cluster architecture is discussed in details in Section  \ref{cmsmonit-k8s}.

\subsection{The CMS monitoring k8s clusters}
\label{cmsmonit-k8s}

As shown in Fig. \ref{cmsmonitinfra}, the CMS monitoring infrastructure is based on two distinct k8s clusters: the NATS cluster that executes the NATS service, and the CMS monitoring cluster to host various services dedicated to  CMS monitoring needs (described in Section \ref{cms-monit-addons}). 

NATS is a simple, secure, and high performance open source messaging system for cloud-native applications. Due to its excellent performance benchmarks, lightweight nature, and easily maintainable infrastructure, it fits well our requirements for real-time monitoring applications. It provides a  basic publisher subscription model for clients written in a variety of programming languages. The NATS cluster is available outside of the CERN firewall, and accessible to all CMS collaborators and services through token-based authentication. The NATS cluster provides the NATS service, which works as a proxy between CMS data providers, such as CMSSW (the CMS software framework), DBS, and WMAgent, and data subscribers located either on the client infrastructure or within the CMS monitoring cluster. In the latter case, we run a series of dedicated NATS subscribers which consume data from the NATS server and feed them into the VictoriaMetrics back-end. 

The CMS monitoring k8s cluster is used to monitor our computing nodes, services and applications. It is deployed in the internal CERN network, and is available on a private network for CMS nodes and services via dedicated firewall rules. The cluster hosts both the Prometheus service that consumes metrics from various exporters running on CMS nodes and services, and the VictoriaMetrics service as long-term storage back-end for the Prometheus server. 
Since both Prometheus and VictoriaMetrics data sources are supported by the MONIT infrastructure, we designed Grafana dashboards to monitor our k8s clusters, the CMSWEB k8s cluster, and various CMS nodes, services and applications. 

A dedicated AlertManager service runs within the CMS monitoring cluster. It is configured to setup various alerts based on key metrics of the monitored applications.
The AlertManager service is tightly integrated into Prometheus and provides a monitoring service to handle all alerts in our monitoring infrastructures. We discuss the details of the alert notification system in Section \ref{alerts}.

The CMS k8s monitoring clusters consists of two nodes with 16 CPU cores and 30 GB RAM each, and hosts 68 individual applications running within the k8s pods. In total only 15\% of the cluster resources are currently used, leaving us room to horizontally scale the provided monitoring services.

\section{CMS monitoring applications}
\label{use-cases}

\begin{figure*}[t]
\centering
\includegraphics[width=0.8\textwidth]{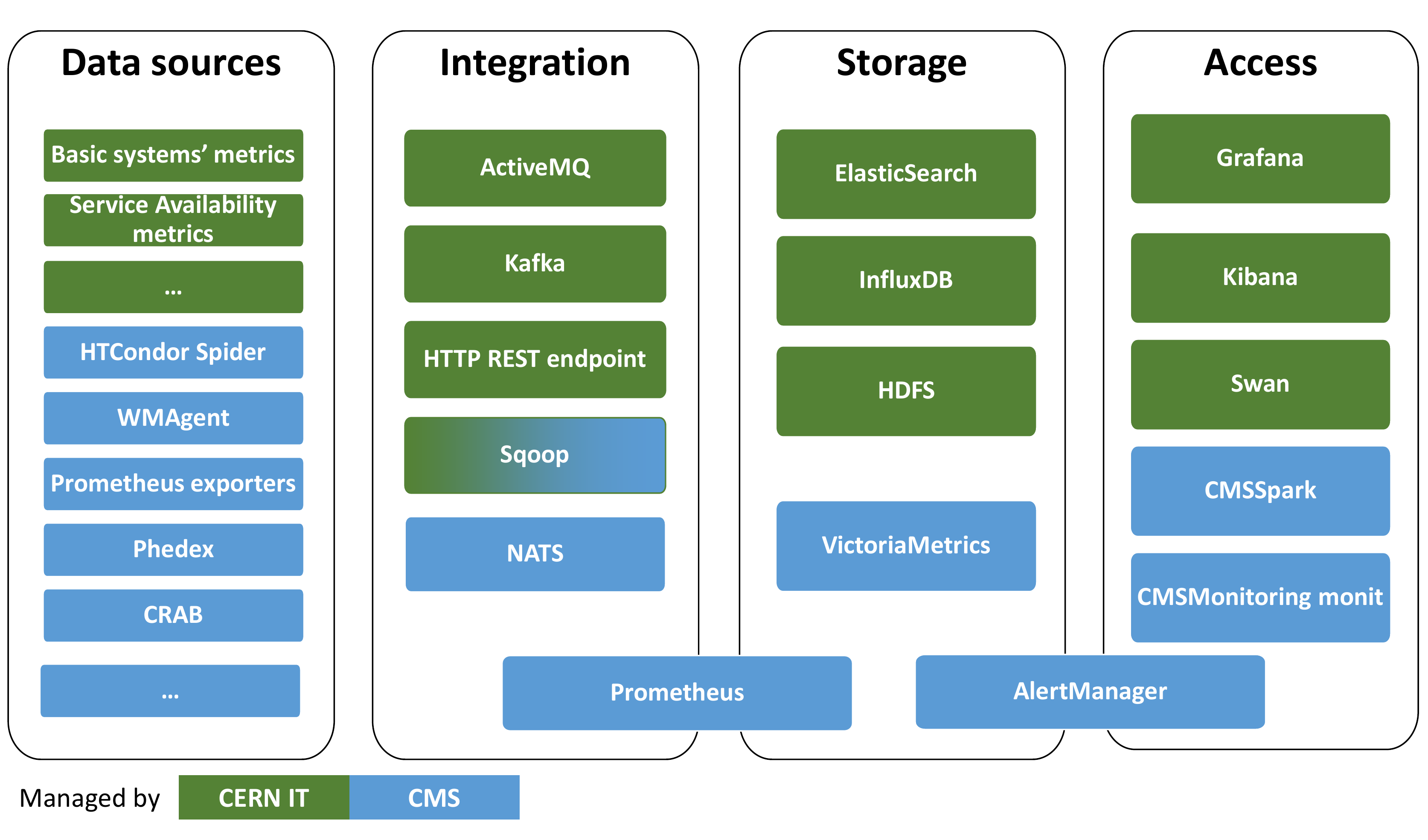}
\caption{Architectural diagram depicting the various components used by the CMS monitoring applications. Components managed by the CERN MONIT team are shown in green, while components managed by CMS are shown in blue. The Sqoop module belongs to both infrastructures since it executes CMS workflows on the HDFS storage system provided by the MONIT infrastructure.}
\label{cms-monit-arq}
\end{figure*}

The architectural diagram of the complete monitoring infrastructure, including all components described in Section \ref{cmsmonit}, is shown in Fig \ref{cms-monit-arq}. In this Section we discuss various monitoring use cases and applications for CMS based on this architecture. We start with an overview of HTCondor job monitoring in Section \ref{cms-prod-mon}. In Section \ref{cms-srv-mon} we provide details of the monitoring applications for CMS services. CMSWEB and k8s monitoring are described in Section \ref{cms-k8s-mon}. Real time monitoring is discussed in Section \ref{real-time-mon}.
Alert handling is presented in Sect. \ref{alerts}. We conclude with a discussion of various command-line tools experts use for monitoring in Section \ref{cms-cli-mon}.

\subsection{HTCondor job monitoring}
\label{cms-prod-mon}

Monitoring central and user activities on the CMS distributed computing resources plays an important role for a variety of stakeholders (computing experts, management, developer and operation teams). This monitoring workflow is based on data collected by the spider service described in Section \ref{cms-monit-experience}. Using the InfluxDB data source we provide a full set of dashboards for CMS collaborators containing several views: the time evolution of jobs in different statuses (completed, running, and pending), utilisation of CPU cores, average CPU efficiency, queue and wall-clock times per core and per CPU. All visualization can be further broken down into various categories such as job types, job input types, computing site where the job was executed. A crucial view of the job monitoring application based on the ES data source allows CMS users to monitor the status of their HTCondor jobs submitted through CRAB. This is an example where additional support indexes needed to be created in order to satisfy all  user requirements for visualization.

\subsection{Services and nodes monitoring}
\label{cms-srv-mon}

Various CMS applications, services, and hardware nodes are monitored using Prometheus, VictoriaMetrics, and AlertManager. Prometheus is a very useful tool to monitor individual services, from a detailed overview of various metrics of a single Linux node to more complex metrics representing service behavior. We use both standard and custom made exporters to scrape service metrics and expose them to our Prometheus servers. The Prometheus service provides a functional query language called PromQL (Prometheus Query Language), that allows users to explore service behaviors. It is integrated with Grafana and supported in the CERN MONIT infrastructure as data source.
VictoriaMetrics provides several enhancements to PromQL, and can be used as a fast storage backend for Prometheus. We leverage this functionality to extend the time retention policy for our metrics.
Alert Manager provides a useful way to write individual alerts based on nodes and service metrics.

All CMS production systems, including CMSWEB services, individual virtual machines (VMs), and k8 clusters, are monitored through these tools. We provide both service specific dashboards and overview dashboards representing the status of nodes, services, databases associated with certain activities in CMS.

\subsection{CMSWEB and k8s monitoring}
\label{cms-k8s-mon}

\begin{figure*}[!t]
\centering
\includegraphics[scale=0.25]{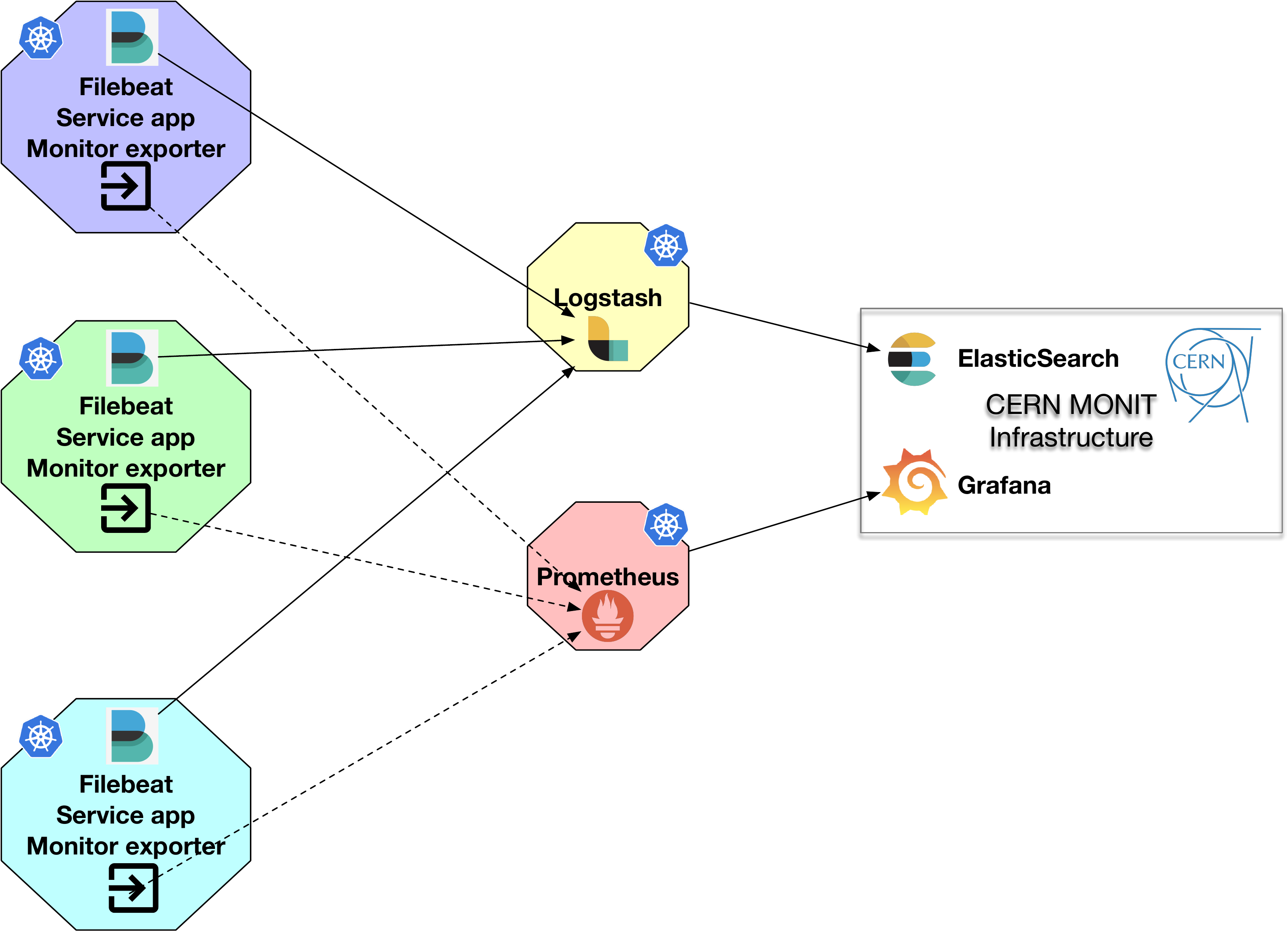}
\caption{The data flow for CMSWEB and k8s logs. Logs are processed and streamed by the Filebeat daemon on each individual k8s pod, and sent to the Logstash service that redirects them to ES. The Prometheus service scrapes service metrics via various exporters and is available as  data source in the CERN MONIT infrastructure. In both cases the information can be visualized with Kibana or Grafana. }
\label{cmsmonit-logs}
\end{figure*}

User activities on the CMSWEB and CMS monitoring k8s clusters are monitored collecting metrics scraped by Prometheus. System and application logs are collected using the Logstash, Filebeat \cite{filebeat}, and ES stack, as shown in Fig. \ref{cmsmonit-logs}.
All logs are streamed into the CERN MONIT and CERN security infrastructures. Selected log metrics are injected into  ES and then visualized in Grafana. For instance, hourly and daily statistics for services on the CMS distributed computing infrastructure are collected. This information is extremely valuable to track the on-going users' activities and the load on our systems, services, databases, and is successfully used to debug various issues with CMS production systems.

In addition, we run independent log scraping and send relevant information, including system and Apache server logs on all production nodes, via Apache Flume \cite{flume} streams to the CERN security infrastructure. This data is securely stored within the CERN security infrastructure, in compliance with the CERN privacy policies, and may be used by the CERN security team to monitor potential hacker activities.

Recently many CMS services were migrated to k8s clusters, including our own CMS monitoring infrastructure. Resources on k8s are monitored through metrics scraped by Prometheus and visualized in Grafana by the Kube eagle \cite{kube-eagle} middleware. The collected metrics show activities on the k8s cluster activities, and are integrated with the AlertManager workflow so that notifications are issued in case of problems with the monitored cluster.

\subsection{Real-time monitoring}
\label{real-time-mon}

Real-time monitoring is useful for several CMS use cases, for example to track failures of workflows executed on the CMS distributed computing resources. CMS users or operators may need to monitor failures at a certain computing center, or for accesses to certain data. The size of the data produced by specific workflows may need to be monitored. Such requests can be made to specific CMS production databases, such as DAS or DBS. These queries create however additional loads to CMS production services, which may have undesirable outcomes such as DOS (Denial of Service). Therefore, we proposed and implemented a (quasi) real-time monitoring service for typical use cases as those above described through NATS (introduced in Section \ref{cms-monit-addons}). 

We integrated DAS publishers into  the CMS production services, and collect the required information in the NATS server. The benefit of using the NATS server is that it works as a proxy between data publishers and subscribers, i.e. it does not store the injected metrics, but guarantees their redirection to any number of subscribers. In other words, it works like a one-way chat messaging system where publishers push the message and any subscriber can receive it right away.  Since the data volume for message exchange can be high, and quite often happening in a burst, we kept the NATS server separated from the CMS Monitoring cluster, and deployed it into a dedicated k8s cluster, as shown in Fig. \ref{cmsmonitinfra}. This allows us to separate the load between the distributed agents (the publishers) and the CMS subscribers. Several subscribers within the CMS monitoring cluster capture the flowing information and redirect it to the VictoriaMetrics backend, so that it can be later used to build Grafana dashboards for a variety of use cases.

\subsection{Alerts}
\label{alerts}

Alerts play an important role in a constantly growing monitoring infrastructure. Grafana provides simple threshold-based alerts that can be set up for any supported data sources. Metrics scraped by Prometheus and VictoriaMetrics  are used to set up more complex rule-based alerts in AlertManager. Alerts can be configured to be sent out to various channels: email notifications, ticketing systems, or communication platform such as Slack \cite{slack}.

However these systems lack the capability to intelligently group alerts, and correlate them with already known system outages. We often found that, while it is desirable to get proper notifications about system misbehavior, human operators cannot cope with a sustained flow of alerts. This typically happens in cases where the same event raises multiple failures and therefore multiple alerts from several systems.

We are currently developing an intelligent alert system that takes into account system outages notifications from several data sources, such as the Site Status Board and several ticketing systems (ServiceNow \cite{snow}, GGUS \cite{ggus}), and overlays them with alerts coming from our metrics. For this purpose we exploit several features of AlertManager: alerts chaining, grouping, and silencing.  

\subsection{Command line tools}
\label{cms-cli-mon}

While the majority of CMS monitoring use cases are covered by the broad spectrum of visualization capabilities of Grafana and Kibana, for some cases Command Line Interface (CLI) tools are required. For instance, some services need to programmatically access the data sources in the MONIT infrastructure, or some services running behind a firewall need to provide terminal based monitoring. In this section we review several CLI tools that we developed in the past months. The degree of adoption in the CMS computing community of such tools strongly depends on their ease of use. We opted to write our tools in the Go programming language since it can provide static executables for all computer architectures supported by CMS. 

The following set of tools are available on the CMS distributed infrastructure through CVMFS \cite{cvmfs}:
\begin{itemize}

\item \textbf{monit}: a tool to query ES and InfluxDB as well as to inject data to ES through a Grafana proxy;
\item \textbf{NATS}: subscriber and publisher CLI tools;
\item \textbf{dbs\_vm}: a tool to query VictoriaMetrics;
\item \textbf{grafterm}\cite{grafterm}: a tool to visualize metrics stored in Prometheus or VictoriaMetrics in a terminal-based user interface.
\end{itemize}

\section{Results}
\label{results}

We present a summary of various measurements (data volumes, data sizes, usage statistics) related to the various CMS monitoring components described in the previous sections. 


Rates of received and sent messages on the ActiveMQ brokers range between 4 KHz and 7.5 KHz. On average, CMS producers send more than 3.5 million messages per hour. Most of the messages are sent by the spider.



In total we maintain forty data sources in ES, twenty-seven in InfluxDB, eighteen in Prometheus, and collect data in more than twenty HDFS locations. The data in ES are  organised in daily indexes corresponding to more than twenty different document types specific to CMS. The total size amounts to more than 20 TB.
The data volume on ES is mostly driven by the HTCondor job data, with a daily index size of about 30 GB, while the other data sources have daily index sizes of 1 GB or less.

On HDFS we currently store around 300 GB per day before compaction. The compaction process takes care of deleting duplicate records and compresses the JSON documents reducing the storage needs for historical data of a factor up to 90\%. After compaction, the CMS data sources account for around 32 GB per day. The largest data source is the HTCondor job monitoring metrics that currently has a size of 22.5 TB.

    
The total data volume in Prometheus and VictoriaMetrics is smaller than that in MONIT, but is gradually growing. Currently we store 30000 active time series for a total of 15.7 billions data points at an ingestion rate of 4.2 kHz. The total number of entries in the inverted index is 11 millions, and the daily time series churn rate is 30000. The total size on disk amounts to 8 GB, with a total index size of 170 MB. The average query range duration is 30 ms.
The Prometheus service currently covers more than one hundred of nodes with 125 exporters, more than three thousand measurements, and provides almost one hundred different alert records and alert rules.


In total, the CMS computing community built more than three hundred Grafana dashboards using all available data sources. On average CMS users are accessing daily more than thirty different dashboards. More than fifty Grafana alert rules are setup to send alerts to about twenty different channels, while in AlertManager we configured fourteen different receivers. Several Spark-based workflows periodically generate a variety of views to monitor specific aspects, for example CMS data access and popularity, and HTCondor job metrics such as CPU efficiency and memory consumption. 


%
%

\section{Summary}
\label{summary}

In the era of distributed computing, the monitoring infrastructure plays an important role to ensure the efficient operation of computing nodes, services such as data management and workflow management, computing clusters, and distributed facilities. We presented a global overview of the CMS monitoring infrastructure, which leverages both the CERN MONIT infrastructure and a variety of additional monitoring services and applications deployed on in-house k8s clusters. We discussed architectural choices, lessons, and presented measurements of various performance metrics and statistics.  

Based on our experience, the choice of open source tools is the key to build scalable and maintainable applications in such a complex heterogeneous environment.  The various tools discussed in a paper address specific functionalities, from data injection to data visualization, and their choice is driven by the need for ease of maintenance and sustainable evolution of the infrastructure. In particular, we show that the CERN MONIT infrastructure is suitable for pushing monitoring data from distributed data providers, while the stack composed by Prometheus, VictoriaMetrics, and AlertManager covers specific needs of CMS internal data services. The adoption of Kubernetes significantly simplifies the deployment of all the tools, and allows to build a scalable infrastructure based on  dynamic resource allocation. 
The chosen visualization solutions, Kibana and Grafana have been proven to be easy-to-use tools for interactive data exploration and production quality dashboards respectively. In conclusion, we are confident that the current monitoring strategy will fulfill CMS expectations and challenges in  the up-coming HL-LHC (High Luminosity LHC) era, where experiments will be required to cope with at least a factor ten higher data rates.

\begin{acknowledgement}

We thank the CERN IT department, and in particular the MONIT team, for the successful collaboration throughout these years. We also thank our CMS colleagues who contributed to the migration effort and helped us building a solid monitoring infrastructure.
This work has received funding from the European Union’s Horizon 2020 research and innovation program under the Marie Skłodowska-Curie grant agreement LHCBIGDATA No. 799062.

\end{acknowledgement}

%
%


\end{document}